# Hybrid Adaptive Fuzzy Extreme Learning Machine for text classification

Ming Li, Peilun Xiao, and Ju Zhang

In traditional ELM and its improved versions suffer from the problems of outliers or noises due to overfitting and imbalance due to distribution. We propose a novel hybrid adaptive fuzzy ELM(HA-FELM), which introduces a fuzzy membership function to the traditional ELM method to deal with the above problems. We define the fuzzy membership function not only basing on the distance between each sample and the center of the class but also the density among samples which based on the quantum harmonic oscillator model. The proposed fuzzy membership function overcomes the shortcoming of the traditional fuzzy membership function and could make itself adjusted according to the specific distribution of different samples adaptively. Experiments show the proposed HA-FELM can produce better performance than SVM, ELM, and RELM in text classification.

*Introduction:* ELM has been applied to text classification research in recent years, because of that ELM has faster learning speed and better generalization performance compared with traditional machine learning method. Xiang guo Zhao et al. proposed a Bagging-ELM method in document classification, they applied the REV and RCC methods to Bagging-ELM successfully and achieved better results than ELM [1]. Li juan Duan used KELM to classify historical patent documents and achieved better results than SVM [2]. Yu haiyan et al. reduced the text feature by information gain, and introduced wavelet into KELM to conduct the emotional classification of Chinese text [3]. Li Yongqiang proposed CPSO-ELM algorithm to select the input weights and bias of hidden nodes in ELM to classify XML documents by optimizing search strategy[4].

The traditional ELM often has an overfitting problem when outliers exist in the training datasets. Fuzzy ELM provides an effective approach to deal with the problem[5], which reduces the effects of outliers by fuzzy membership functions. In this paper, we propose a novel Hybrid Adaptive Fuzzy ELM (HA-FELM). To construct a suitable fuzzy membership function, we should not only consider the distance between the samples and the center of the class but also consider the density of the samples. In this paper, we design a hybrid adaptive fuzzy membership function based on distance and density. The traditional fuzzy density membership depends on the density of k samples that are the nearest neighbor to the sample, which is difficult to reflect the sample distribution. We utilize the advantage of clustering algorithm based on quantum harmonic oscillator model(CA-QHO), which can obtain the actual clustering number without prior knowledge, and automatically cluster the dataset into classes according to the distribution of samples. We introduce this clustering algorithm into fuzzy density membership function, so the density based membership function can be calculated according to the different distribution of samples adaptively.

*Fuzzy extreme learning machine:* FELM introduces a fuzzy matrix S into ELM to reduce the effect of outliers or noises in the learning of a classification model. FELM assigns different training samples with different fuzzy membership values, such that more important samples are assigned with higher membership values while less important samples are assigned with lower membership values.

Given Set $\{(x_i, y_i) | x_i \in R^n, y_i \in R^m, i = 1, 2, \ldots, N\}$, wh-ere $x_i$ is the sample, $y_i$ is the class label, $g(x)$ is the hidden layer activation function, $s_i$ is a fuzzy membership ($0 < s_i \leq 1$), FELM mathematical model can be expressed as:

$$Minimize: \frac{1}{2}\|\beta\|^2 + C\frac{1}{2}\sum_{i=1}^{N} s_i \|\xi_i\|^2$$
$$Subject\ to: \mathbf{h}(x_i) = \mathbf{t}_i^T - \xi_i^T \quad i = 1, \ldots, N \quad (1)$$

Derived by KKT conditions:

$$\beta = \mathbf{H}^T \left(\frac{\mathbf{S}}{\mathbf{C}} + \mathbf{H}\mathbf{H}^T\right)^{-1} \mathbf{T} \quad (2)$$

fuzzy matrix $\mathbf{S}$ is diagonal matrix, each element in $S \in R^{N \times N}$ is $1/s_i$. Given a testing sample $x$, the output vector of WELM is $f(x) = h(x)\beta$, the label of $x$ can be achieved by:

$$label(x) = \arg\max_i f_i(x), i \in \{1, \ldots, m\} \quad (3)$$

*Hybrid Adaptive Fuzzy Membership Function:*
1) Fuzzy membership function based on distance

We determine the fuzzy membership value of each sample according to the distance between the sample and the class center. The sample is closer to the center, the membership value is greater; otherwise, the membership value is smaller. [6] gives several commonly used distance-based fuzzy membership functions. Given Set $\{x_i | x_i \in R^n, i = 1, \ldots, n\}$, where $x_i$ is the sample, we define the fuzzy membership function based on distance as follows:

$$\mu(x_i) = e^{1 - \frac{d_i^{cen}}{d_{\max} + \theta}} \quad (4)$$

Where $d_{\max} = \max_{x_i} \|x_i - \overline{x}\|$, $d_i^{cen} = \|x_i - \overline{x}\|$, $\overline{x} = \sum_{i=1}^{n} x_i$, $\theta$ is a small constant.

2) Fuzzy membership function based on density

The fuzzy membership function based on distance can not reflect the importance of samples accurately, and leads to deviation of classification results, if we only consider the distance factor. We have to think about distance factor and also density factor, when we compute fuzzy membership value. Given Set $\{x_i | x_i \in R^n, i = 1, \ldots, n\}$, where $x_i$ is the sample, we define the fuzzy membership function based on density as follows:

$$\omega(x_i) = \sum_{i=1}^{n} e^{-d_{ij}} \quad (5)$$

Where $d_{ij}$ is the distance of $x_i$ to its $k$ nearest neighbor $x_j (i = 1, \ldots, n)$, the parameter $k$ is user-specified constant in advance. Setting parameter k in advance but not The parameter k is set not according to different sample distributions but in advance, which could lead to density measurement deviation. We will solve this problem in the CA-QHO algorithm later.

Combining with formula (4) and (5), we propose a hybrid adaptive fuzzy membership function :

$$s(x_i) = \alpha \cdot \mu(x_i) + (1 - \alpha)\omega(x_i) \quad (6)$$

*Clustering algorithm based on quantum harmonic oscillator model(CA-QHO):* The physical change of the quantum harmonic oscillator to the ground state is very similar to the cluster centers in the clustering algorithm. Analogous to the physical process of the quantum harmonic oscillator, CA-QHO algorithm is depicted as follow:
(1) Divide the dataset into n two-dimensional grids;
(2) Project the samples in the dataset onto the grid;
(3) Select m samples $(x_{i0}, y_{i0})$ as the initial center point randomly; calculate the density $A_{i0}$ of the grid which the sample $(x_{i0}, y_{i0})$ is belong to;
(4) For $i = 1 : m$

  (q) Set $x_{i0}$ as the mean value of Gaussian random sampling on the x-axis, set $\sigma_x$ as the standard deviation of the x-axis,

construct a Gaussian function $N(x_{i0}, \sigma_x^2)$ on the x-axis, form a Gaussian sampling area, record the $x_{i1}$ of the maximum density value of grids in this area;

(b) Conduct the same operation as the x-axis on the y-axis, get the $y_{i1}$;

(c) Record the grid density $A_{i1}$ corresponding to sample $(x_{i1}, y_{i1})$;

(d) if $A_{i0} - A_{i1} \geq 0$

$x_{i0} = x_{i1}$;

($x_{i1}$ is as the new mean value of Gaussian random sampling on the x-axis);

$y_{i0} = y_{i1}$;

($y_{i1}$ is as the new mean value of Gaussian random sampling on the y-axis);

Go to (q);

(e) Record ith value of $(x_{i0}, y_{i0})$, which is the ith centres of m clusters;

*Text representation:* The high dimensional features will increase the computational burden of ELM. Compared with the traditional method of extracting feature vector, word-vector text representation has better feature expression ability and can avoid the dimension disaster of the feature vector effectively[7]. The word vector model used in this paper is Skip-gram model. \We use the Word2vec tool to generate the word vector first, then generate the document vector by averaging pooling the word vector. We use $c_{i,j}$ to denote the word vector of the $j$th word in the document $i$, the document vector $v_i$, can be calculated as: $v_i = (1/J_i) \sum_{j=1}^{J_i} c_{i,j}$, where $J_i$ is the number of words in the document.

*Results and discussion:* All experiments were conducted with 3.6GHz CPU and 4GB Memory. We used the word2vec tool provided by Google used to train the skip-gram word vector. We conducted text classification experiments on SVM which are build by the scikit-learn tool. The related all algorithms based on ELM are implemented in python. We take the mf1 and MF1 of document vector 300 dimensions as the experimental observation; choose the RBF function as the activation function of the hidden node. The super parameter c and L of ELM are selected by grid searching method. $\{10^0, 10^{-1}, ..., 10^{-8}\}$ is the search range for c values, $\{100, 200, ...1000\}$ is the search scope of L, $\theta$ is set 0.001, $\alpha$ is set 0.7.

Experiments are conducted on imbalanced datasets (Reuters52 and WebKB) and balanced dataset(20Newsgroups) for evaluating the performance of our proposed approach. The results are compared with SVM, ELM, RELM, DI-FELM, DE-FELM and HA-FELM in text classification. DI-FELM refers to FELM based on distance-based fuzzy membership function. DD-FELM refers to FELM based on density-based fuzzy membership function. HA-FELM refers to FELM based on hybrid adaptive fuzzy membership function.

From Table1, it can be seen that mf1 and MF1 of HA-FELM are higher than all the other methods on the imbalanced dataset Reuters52 and WebKB; and HA-FELM is the second in the list of classification comparision on the balanced dataset 20Newsgroups. The performance of HA-FELM is more obvious on the imbalanced dataset Reuters52 and WebKB. HA-FELM shows stable performance, indicating that this method has good generalization performance.

**Table 1:** Comparision of Classification Results on 20Newsgroups, R52 and WebKB

| method | 20News | | R52 | | WebKB | |
|---|---|---|---|---|---|---|
| | mf1 | MF1 | mf1 | MF1 | mf1 | MF1 |
| SVM | 0.766 | 0.755 | 0.923 | 0.592 | 0.873 | 0.857 |
| ELM | 0.770 | 0.756 | 0.925 | 0.592 | 0.875 | 0.862 |
| RELM | 0.783 | 0.770 | 0.927 | 0.527 | 0.877 | 0.862 |
| DI-FELM | 0.785 | 0.772 | 0.922 | 0.536 | 0.879 | 0.866 |
| DE-FELM | 0.782 | 0.770 | 0.926 | 0.527 | 0.875 | 0.858 |
| HA-FELM | 0.783 | 0.771 | 0.930 | 0.544 | 0.88 | 0.867 |

*Conclusion:* In this paper, we propose a hybrid adaptive fuzzy extreme learning machine model and apply it in text classification, which could overcome the problems of noises and imbalance. We assign low fuzzy-membership values for training samples in large categories or contaminated by noises or imbalance. We note the optimal membership function based on sample distribution, so we design the CA-QHO algorithm and apply it to density-based fuzzy membership function. In general, HA-FELM achieves higher predictive accuracy compared to the other five algorithms.

*Acknowledgments:* This research was supported by National Natural Science Foundation of China(No.61672488), Ministry of science and Technology "Key technologies of educational cloud"(No.2013BAH72B01), Chongqing Science & Technology Commission(No.cstc2015shms-ztzx10005), Clinical Novel Technology Funding of Southwest Hospital(No.SWH2016ZDCX1008 ).

Ming Li and Ju Zhang(High performance computing application R&D Center, Chongqing Institute of Green and Intelligent Technology, Chinese Academy of Sciences, Chongqing ,400714, China)
E-mail:liming.cigit.cas@gmail.com
Peilun Xiao(Center for Speech and Language Technology,Research Institute of Information Technology, Tsinghua University,Beijing, 100084,China)
Ming Li(also with University of Chinese Academy of Sciences, Beijing,100049, China; Center for Speech and Language Technology,Research Institute of Information Technology, Tsinghua University, Beijing, 100084, China)